\documentclass[a4paper, 12pt]{article}

\usepackage[utf8]{inputenc}
\usepackage[T1]{fontenc}
\usepackage[english]{babel}

\usepackage{geometry}
\usepackage{amsmath}
\usepackage{amssymb}
\usepackage{amsfonts}
\usepackage{mathtools}

\usepackage{graphicx}
\usepackage{float} % Per posizionare le figure in modo preciso
\usepackage[font=small, labelfont=bf]{caption} % Per personalizzare le didascalie

\usepackage{array}
\usepackage{booktabs} % Migliora l'aspetto delle tabelle
\usepackage{tabularx} % Tabelle a larghezza variabile

\usepackage{xcolor}

\usepackage{hyperref}
\hypersetup{
    colorlinks=true,
    linkcolor=blue,
    citecolor=blue,
    filecolor=magenta,
    urlcolor=blue
}

\usepackage{natbib}
\usepackage{algorithm}
\usepackage{algorithmic}

\usepackage{enumitem}

\usepackage{lmodern}

\usepackage{bm,times,graphicx,latexsym,amsfonts}
\usepackage[english]{babel}
\usepackage{amssymb, amsmath}
\usepackage{mathrsfs}
\usepackage{authblk}

\newtheorem{definition}{Definition} % Define a new environment for definitions

\begin{document}

\title{\textbf{Novel Fuzzy Centrality Measures in Vague Social Networks}}
\author[1]{Annamaria Porreca, \thanks{annamaria.porreca@unimercatorum.it, ORCID: \href{https://orcid.org/0000-0003-3278-1561}{0000-0003-3278-1561}}}
\author[2]{Fabrizio Maturo, \thanks{fabrizio.maturo@unimercatorum.it, ORCID: \href{https://orcid.org/0000-0002-2362-4970}{0000-0002-2362-4970}}}
\author[3]{Viviana Ventre, \thanks{viviana.ventre@unicampania.it, ORCID: \href{https://orcid.org/0000-0001-5314-5770}{0000-0001-5314-5770}}}

\affil[1]{Department of Economics, Statistics and Business; Faculty of Economics and Law; Universitas Mercatorum, Rome, Italy}
\affil[2]{Department of Economics, Statistics and Business, Faculty of Technological and Innovation Sciences, Universitas Mercatorum, Rome, Italy}
\affil[3]{University of Campania Luigi Vanvitelli, Mathematics and Physics, Caserta, Italy}

\date{}

\maketitle

\begin{abstract}
\noindent Social network analysis (SNA) helps us understand the relationships and interactions between individuals, groups, organizations, or other social entities. In the literature, ties are generally considered binary or weighted based on their strength. Nonetheless, when the actors are individuals, these relationships are often imprecise, and identifying them with simple scalars leads to information loss. Indeed, social relationships are often vague in real life, and although previous research has proposed the use of fuzzy networks, these are typically characterized by crisp ties.
The use of weighted links does not align with the original philosophy of fuzzy logic, which instead aims to preserve the vagueness inherent in human language and real life.
For this reason, this paper proposes a generalization of the so-called Fuzzy Social Network Analysis (FSNA) to the context of imprecise relationships among actors.
Dealing with imprecise ties and introducing fuzziness in the definition of relationships requires an extension of social network analysis, defining ties as fuzzy numbers instead of crisp values and extending classical centrality indices to fuzzy centrality indexes.
The article presents the theory and application of real data collected through a fascinating mouse-tracking technique to study the fuzzy relationships in a collaboration network among the members of a university department.\\
\textbf{Keywords}: Fuzzy Social Network,  Fuzzy Centrality Indices,  Vague Relationships,  Uncertainty, Fuzzy Numbers,  University Department Collaboration Network.
\end{abstract}

\newpage

\section{Introduction}

Social network analysis (SNA) helps us understand the relationships and interactions between individuals, groups, organisations, or other social entities \citep{hanneman2011,otte2002, opsahl2010, wasserman1994, Landherr_2010}. Each type of social structure can be represented with a graph, in which the social entities are represented as nodes or vertices in a network, and their relationships are represented as edges or ties. SNA examines relationships between entities and how information, resources, or influence flows through them. It identifies patterns, structures, and dynamics of social systems.
Ties in a network are generally binary based on the presence or absence of the link or weighted based on their strength. Each connection can be identified by a number in $[0,1]$ or $\{0,1\}$, respectively. 

Nonetheless, when the actors of a social network are individuals, the relationships are very complex and measuring them using a single number leads to a loss of information. In other words, relationships are inherently imprecise in real life, and thus, social network analysis should consider vagueness.  
Consider a social network like Facebook, where nodes represent users and links denote friendships. However, not all friendships carry the same meaning: some are more interactive, while others are more distant. This mirrors the complexity of real-life relationships, which are rarely clear-cut but often vague. The concept of "friendship" or "collaboration" is inherently ambiguous, varying with context, and highlighting the uncertainty present in both online interactions and real-life connections.
The concepts of friendship, close friendship, collaborations, or acquaintances are inherently vague, and representing them with a single number sacrifices information for the sake of simplicity.

The most natural way to consider the problem mentioned above is to introduce a fuzzy-based social network analysis. 
In the literature, fuzzy network analysis has mainly been studied by introducing weighted ties within a network \citep[see, e.g.,][]{hu2015centrality, brunelli2009fuzzy}. However, weighting the links between nodes has nothing to do with fuzzy logic, and using the term ``fuzzy'' raises some doubts because it suggests an approach that considers the vagueness of relationships, which only in this latter case would be in line with the basic idea of fuzzy logic that begins and thrives where opposites coexist. Indeed, fuzzy logic deals with situations where A and non-A coexist simultaneously with a certain degree of truth \citep{Zadeh1, Zadeh2, Zadeh3, yager1980choosing}. Hence, translating vague concepts into crisp values necessarily results in a loss of information regarding the inherently imprecise nature of these expressions. In other words, assigning a link a value in $[0,1]$ or $\{0,1\}$ can sometimes be a simplification that sacrifices information for the sake of convenience.

Dealing with imprecise ties and introducing fuzziness in the definition of relationships needs an extension of classical social network analysis to fuzzy numbers instead of crisp values. The mathematical formalisation of this generalisation requires rewriting the classical centrality indices into fuzzy centrality indices involving fuzzy relationships. The main motivation of this study is that, in many contexts, relationships between nodes in a social network should be characterised by functions rather than scalars to preserve the information about the vagueness of ties. 
The article proposes new fascinating Fuzzy Centrality Measures (FCMs) for Fuzzy Social Networks (FSNs) and show an application to real data. The dataset adopted in this study is collected through a mouse tracking technique to capture the vagueness of the relationships among the members of a University department. 
The best results of this research and the main advantages of our proposal are the creation of social networks based on vague ties and centrality indices preserving information about their imprecision, inherent in human nature, human language, and interpersonal relationships. 

The first part of the work introduces the preliminaries necessary for formally understanding the proposed indices. Subsequently, based on the idea mentioned above,  new centrality indices are proposed: fuzzy degree centrality, fuzzy out-degree centrality index, total fuzzy degree centrality index, fuzzy betweenness centrality, fuzzy closeness centrality, and fuzzy out-closeness centrality. The application Section proposes a fascinating investigation based on a University department collaboration network. The paper ends with a discussion and conclusions.

\section{Background on Centrality Indices in Social Networks}

Centrality indices are essential tools in social network analysis, capturing the importance and influence of nodes within a network, and have been a topic of research for many decades.
\citet{bavelas1948mathematical} pioneered centrality measures in the paper ``\textit{A Mathematical Model for Group Structures}'' presenting one of the earliest formal approaches to understanding social networks and group structures through mathematical modeling. He explores how the arrangement of communication pathways within a group affects the group's efficiency, centralization, and overall functioning.
\citet{Katz_1953} introduced the concept of Katz centrality, which extends the idea of degree centrality by considering the influence of both direct and indirect connections. Katz centrality assigns more weight to closer (direct) connections while also accounting for longer paths in the network.
The concept of centrality has been explored in various forms, with foundational contributions such as the general framework for centrality indices proposed by \citet{sabidussi1966centrality}, which provided a unified approach to understanding node importance within a network.
Closeness centrality was introduced by Linton C. Freeman in his seminal 1979 paper titled ``\textit{Centrality in Social Networks: Conceptual Clarification}'' \citep{freeman2002centrality}. In this paper, Freeman formalized various centrality measures, including closeness centrality, which measures how close a node is to all other nodes in a network. The concept was developed to capture the idea that a node is more central if it can quickly interact with all other nodes, reflecting its potential to communicate or spread information efficiently within the network.
In their book ``\textit{Social Network Analysis: Methods and Applications}'', \citet{wasserman1994} provided a comprehensive treatment of various centrality measures and their applications in social network analysis. Their work has been instrumental in standardizing centrality concepts and methods in the field.
\citet{opsahl2010} proposed generalizations of degree and shortest-path centralities to weighted networks, providing a more comprehensive view of centrality in networks where connections have varying strengths. These measures take into account both the number and weight of connections, offering a more accurate reflection of a node's importance. These indices remain fundamental in understanding the roles and significance of nodes within complex networks.
In their critical review, Landherr et al. assessed various centrality measures, highlighting the strengths and limitations of each within different network contexts. This work is valuable for understanding how different centrality measures can be applied depending on the specific characteristics of a network \citep{Landherr_2010}.
\citet{blue2002unified} contributed to the application of fuzzy logic in graph theory and, by extension, social networks. Their work involves developing fuzzy graph models that can be applied to social network analysis, allowing for the representation of uncertainty and partial membership in social ties.
\citet{brunelli2009fuzzy} discussed a fuzzy approach to social network analysis, including the development of fuzzy centrality indices. Their approach considered the fuzziness of social ties and how this impacts the calculation of centrality within networks. However, in their framework, the ties between nodes are represented by crisp membership values, which indicate the degree of membership or connection strength within the fuzzy context.

\section{Preliminaries}

\subsection{Fuzzy Numbers}

A fuzzy set \emph{A} is defined by a set of ordered pairs and a binary relation as follows:

\begin{equation}
\emph{A} = \{(x, \mu_A(x)) \mid x \in A, \mu_A(x) \in [0,1]\}
\end{equation}

\noindent where $\mu_A(x)$ is a function, known as the membership function, that specifies the degree to which any element $x$ in $A$ belongs to the fuzzy set \emph{A}. This implies a gradual transition from membership to non-membership \citep{Zadeh2}. Formally, a fuzzy set \emph{A} is defined by its characteristic function:

\begin{equation}
\mu_A:X\rightarrow [0,1]
\end{equation}

\noindent being $X$ the universe of definition.

A fuzzy number $\mu(x)$ is a special case of a fuzzy set; in fact it can be defined as a fuzzy set, defined on real number, with a normal and convex membership function, such that there exists at least one point where the membership function takes the value ``one''; it is a very useful tool for working with imprecise numerical quantities \citep{Zadeh1, maturo2020fuzzy, maturo2017fuzzy, maturo2016dealing}.
A fuzzy number is a function having as domain the set of real numbers and with values in $[0, 1]$:

\begin{equation}
\mu: \mathbb{R} \rightarrow [0,1]
\end{equation}

\noindent such that the following characteristics apply. 1) Bounded support: there are two real numbers \textit{a} and \textit{b}, with $a\leq b$, called the endpoints of $\mu$, such that     $\mu(x) = 0$  for  $x\not\in [a,b]$ and 	$\mu(x) > 0$ for $x\in (a,b)$; 2) Normality: there are two real numbers $c$ and $d$, with $a \leq  c \leq d \leq b$ such that $\mu(x) = 1$ if and only if $x\in[c,d]$; 3) Convexity: $\mu(x)$ is a function increasing in the interval $[a, c]$ and decreasing in the interval $[d, b]$; 4) Compactness: for every $\alpha\in (0,1)$, the set \{$x\in \mathbb{R}: \mu(x)=\alpha$\} is a closed interval. The set of the real numbers $x$ such that $\mu(x) > 0$ is said the support of the fuzzy number, and the interval $[c, d]$ is the core or central part. The intervals $[a, c)$ and $(d, b]$ are, respectively, the left part and the right part.
The real numbers ${\mu}_L=c-a$, ${\mu}_C=d-c$, and ${\mu}_R=b-d$ are the left, middle, and right spreads, respectively. Their sum ${\mu}_T=b-a$ is the total spread of the fuzzy number \citep{Klir}.

\begin{figure}[htbp]
\centering
\includegraphics[width=6cm]{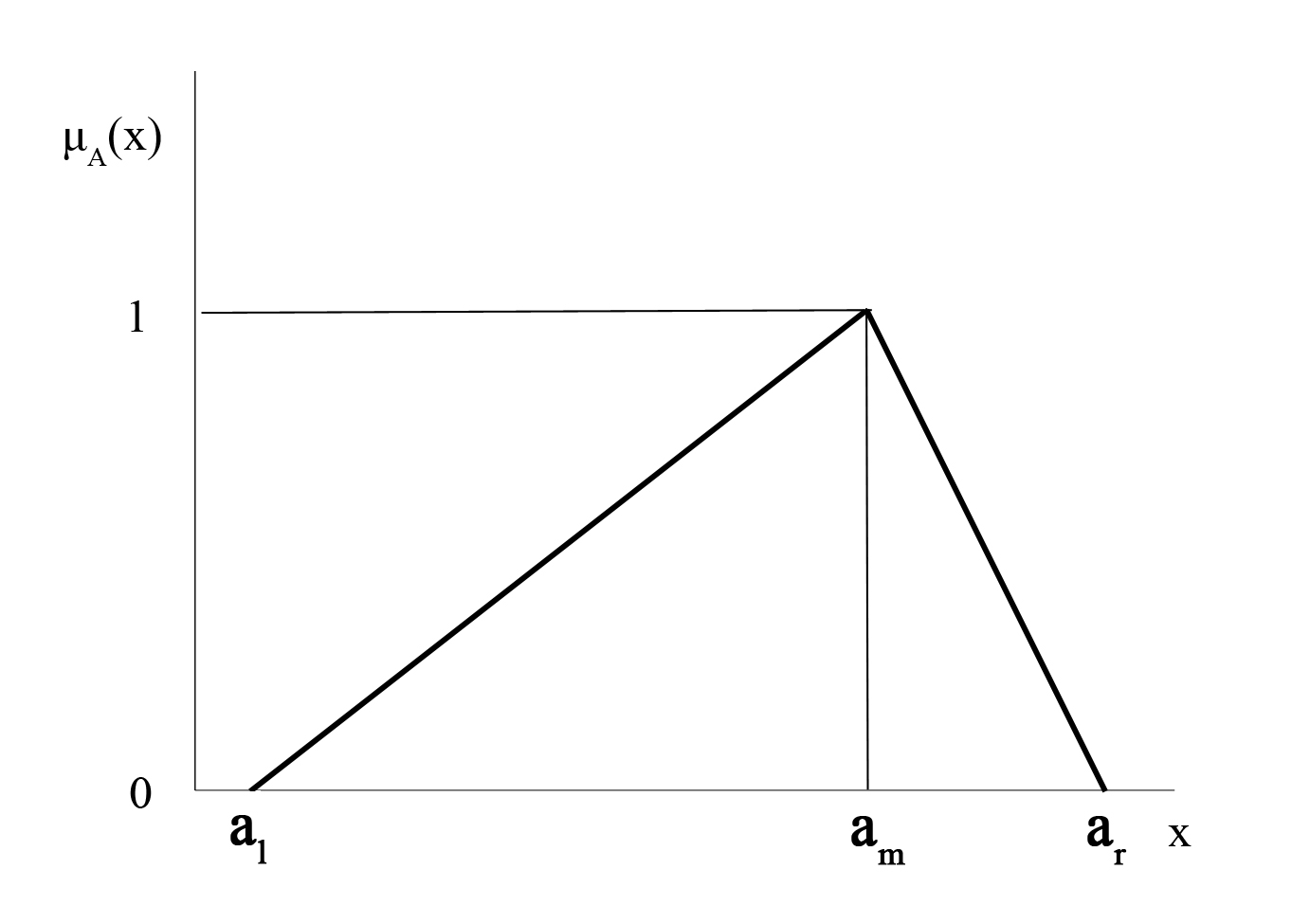}
\caption{Triangular fuzzy number (TFN).}\label{triangular}
\end{figure}

The choice of the type of fuzzy numbers influences the characteristics of the processes of fuzzification and defuzzification of the outputs and the inputs \citep{Zadeh2}.
The primary membership functions used to represent fuzzy variables include triangular (TFN), trapezoidal, and bell-shaped functions. In this work, we focus specifically on TFNs. A TFN $A$ is defined by the following membership function:

\begin{equation}
\mu_A(x)=
\begin{cases}
    \frac{x-a_l}{a_m-a_l} & \text{for} \quad a_l\leq x\leq a_m\\ 	
    \frac{x-a_r}{a_m-a_r} & \text{for} \quad a_m\leq x\leq a_r\\ 	
    \ 0 & \text{otherwise}.
    \end{cases}
\end{equation}

\noindent where $[a_l,a_r]$ is the support and $a_m$ is the core. As illustrated in Fig.\ref{triangular} $a_l$ and $a_r$ are respectively the left and right endpoints, while $a_m$ is the point where the membership function is equal to one.
However, a TFN is often indicated using a simpler notation as follows: 
$A=(a_l,a_m,a_r)$.

\subsection{Basic Pseudo-Operations on Triangular Fuzzy Numbers}

In the context of fuzzy logic, it is often necessary to extend classical arithmetic operations to fuzzy numbers. These extended operations are referred to as \textit{pseudo-operations}. 
Specifically, pseudo-operations allow us to perform arithmetic in a fuzzy environment\citep{simo2018fuzzy}.
Given two triangular fuzzy numbers \(\tilde{A} = (a_l, a_m, a_r)\) and \(\tilde{B} = (b_l, b_m, b_r)\), the following pseudo-operations are defined:

\begin{enumerate}
    \item Addition (\(\oplus\)): 
    \begin{equation}
    \tilde{A} \oplus \tilde{B} = (a_l + b_l, a_m + b_m, a_r + b_r)
    \label{eq:addition}
    \end{equation}
    
    \item Multiplication (\(\otimes\)):    
    \begin{equation}
    \tilde{A} \otimes \tilde{B} = (a_l \cdot b_l, a_m \cdot b_m, a_r \cdot b_r)
    \label{eq:multiplication}
    \end{equation}
    
    \item Scalar Multiplication (\(\odot\)):    
    \begin{equation}
    w \odot \tilde{A} = (w \cdot a_l, w \cdot a_m, w \cdot a_r)
    \label{eq:scalar_multiplication}
    \end{equation}
\end{enumerate}

\subsection{Fuzzy Triangular Ordered Weighted Arithmetic (FTOWA)}

The Fuzzy Triangular Ordered Weighted Arithmetic (FTOWA) Operator is an aggregation operator specifically designed for triangular fuzzy numbers. The FTOWA operator aggregates a set of triangular fuzzy numbers by considering their ordered weights in a fuzzy environment\citep{simo2018fuzzy}. It is defined as:

\begin{equation}
 \text{FTOWA}(\tilde{A}_1, \tilde{A}_2, \dots, \tilde{A}_n) = \bigoplus_{i=1}^n w_i \odot \tilde{A}_{(i)}
\label{FTOWA}
 \end{equation}

\noindent where  $\tilde{A}_{(i)} = (a_{(i)l}, a_{(i)m}, a_{(i)r})$ is the $i$-th largest TFN in the ordered set, and $w_i$ is the weight associated with the \(i\)-th ordered fuzzy number.

The FTOWA operator includes various specific cases depending on the choice of weights \(w_i\):
\begin{enumerate}
    \item Fuzzy Minimum: If \(w = (1, 0, \dots, 0)\), then FTOWA gives \(\tilde{A}_{(1)}\), the smallest triangular fuzzy number.
    \item Fuzzy Maximum: If \(w = (0, \dots, 0, 1)\), then FTOWA gives \(\tilde{A}_{(n)}\), the largest triangular fuzzy number.
    \item Fuzzy Arithmetic Average: If \(w_i = \frac{1}{n}\) for all \(i\), then FTOWA reduces to the arithmetic average of the triangular fuzzy numbers.
\end{enumerate}

The FTOWA operator preserves essential properties like monotonicity, continuity, and idempotency, making it a versatile tool in decision-making processes where the information is represented as triangular fuzzy numbers.

\subsection{Fuzzy Graphs}

\citet{rosenfeld1975fuzzy} introduced fuzzy graphs by applying fuzzy set theory, where both vertices and edges are characterized by membership values in 
$[0,1]$. Formally:

\begin{definition}[Rosenfeld \citep{rosenfeld1975fuzzy}]
A fuzzy graph $G = (\sigma, \mu)$ consists of functions $\sigma: S \rightarrow [0,1]$ and $\mu: S \times S \rightarrow [0,1]$, where for all $x, y \in S$, $\mu(x, y) \leq \min{\sigma(x), \sigma(y)}$.
\end{definition}

\citet{yeh1975fuzzy} later extended this concept for clustering analysis, allowing both vertices and edges to be fuzzy:

\begin{definition}[Yeh and Bang \citep{yeh1975fuzzy}]
A fuzzy graph $G = (V, R)$ consists of a set of vertices $V$ and a fuzzy relation $R$ on $V$, where edges have a membership function $\mu_{R}: V \times V \rightarrow [0,1]$.
\end{definition}

Subsequently, \citet{blue1997applications,blue2002unified} categorized fuzzy graphs into types, such as fuzzy sets of crisp graphs, crisp vertices and edges with fuzzy connectivity, fuzzy vertices with crisp edges, and crisp graphs with fuzzy weights. These variations reflect different levels of fuzziness in graph structures, from fully crisp to various combinations of fuzzy elements.

In this paper, we focus specifically on the case where the ties (or edges) in the graph are fuzzy, while the actors (or vertices) are crisp. This approach allows us to model the uncertainty or varying strength of relationships between clearly defined actors, reflecting more realistic scenarios in social network analysis where connections may not be strictly binary but can vary in intensity or reliability.
We propose the use of Fuzzy Graphs (FG) as a generalization of Weighted Graphs (WG). A WG is one in which edges are associated with weights. A WG can be represented as \( G(V, E, W) \), where \( W \) represents the weights associated with each edge (\(|W| = |E|\)). For an adjacency matrix representation, instead of \( 0 \) and \( 1 \), we can use the weight associated with the edge. This saves space by combining \( E \) and \( W \) into one adjacency matrix \( A \), assuming an edge exists between \( v_i \) and \( v_j \) if and only if \( W_{i,j} \neq 0 \). Thus, FGs are considered as a generalization of WG, where edges are associated with fuzzy numbers. 

\begin{definition}
An undirected fuzzy social network is defined as a fuzzy relational structure \( \widetilde{G}_{un} = (V, E, \widetilde{A}_{un}) \), where \( V = \{v_{1}, v_{2}, \ldots, v_{n}\} \) is a non-empty set of actors or nodes, and
\[
\widetilde{A}_{un} = \left(
\begin{array}{ccc}
\widetilde{A_{11}} & \cdots & \widetilde{A_{1n}} \\
\vdots & \ddots & \vdots \\
\widetilde{A_{n1}} & \cdots & \widetilde{A_{nn}}
\end{array}
\right)
\]
is an undirected fuzzy relation on \( V \).
\label{defFSN}
\end{definition}

Many fuzzy relations are directional. A fuzzy relation is directional if the ties are oriented from one actor to another. Thus, \citet{hu2015centrality} defined the directed fuzzy social network (DFSN) as follows:

\begin{definition}
A directed fuzzy social network is defined as a fuzzy relational structure \( \widetilde{G}_{dir} = (V, E, \widetilde{A}_{dir}) \), where \( V = \{v_{1}, v_{2}, \ldots, v_{n}\} \) is a non-empty set of actors or nodes, and \( \widetilde{A}_{dir} = \left(\begin{array}{ccc}\widetilde{A_{11}} & \cdots & \widetilde{A_{1n}} \\ \vdots & \ddots & \vdots \\ \widetilde{A_{n1}} & \cdots & \widetilde{A_{nn}}\end{array}\right) \) is an undirected fuzzy relation on \( V \).
\label{defdirected}
\end{definition}

Fuzzy Social Network (FSN) includes Undirected Fuzzy Social Network (UFSN) and Directed Fuzzy Social Network (DFSN). The significant difference between DFSN and UFSN is that a directed fuzzy relation is considered. According to Definition \ref{defFSN}, \( \widetilde{A}_{ij} \) is equal to \( \widetilde{A_{ji}} \) in an undirected fuzzy social network. However, \( \widetilde{A}_{ij} \) is not always equal to \( \widetilde{A}_{ji} \) in a directed fuzzy social network. In Definition \ref{defdirected}, \( \widetilde{A}_{dir} \) is called the directed fuzzy adjacency matrix of \( \widetilde{G}_{dir} \):
\[
\mu_{\widetilde{A}_{dir}} = \left(\begin{array}{ccc}\mu\left(\widetilde{A_{11}}\right) & \cdots & \mu\left(\widetilde{A_{1n}}\right) \\ \vdots & \ddots & \vdots \\ \mu\left(\widetilde{A_{n1}}\right) & \cdots & \mu\left(\widetilde{A_{nn}}\right)\end{array}\right)
\]
where, in this case, we have an asymmetrical membership function. The related concepts of UFSNs can also be used in DFSNs, but the direction of the fuzzy relation between actors must be considered.

\begin{definition}
Suppose \(v_0 A_1 v_1 A_2 v_2 \dots A_k v_k\) is a path from node \(v_0\) to \(v_k\) in a directed fuzzy social network (denoted as \(\widetilde{G}_{\text{dir}}\)).
The directed fuzzy intensity of this path, denoted as \(\tilde{s}_d(\tilde{\omega})\), is calculated as the minimum membership function value among all the edges in the path:
    \[
    \tilde{s}_d(\tilde{\omega}) = \bigwedge_{i=1}^{k} \mu(A_i)
    \]
\end{definition}

\noindent where \(\mu(e_i)\) represents the membership function associated with the edge \(e_i\) in a fuzzy graph or network.

\begin{definition}
If there are \(n\) paths from node \(u\) to node \(v\) in the directed fuzzy social network \(\widetilde{G}_{\text{dir}}\), the directed fuzzy connected intensity from \(u\) to \(v\) is defined as:
    \[
    \widetilde{s}_d(u, v) = \bigvee_{k=1}^{n} \widetilde{s}_d(\widetilde{\omega}_k)
    \]
    where \(\widetilde{\omega}_k\) represents each path from \(u\) to \(v\), and \(\bigvee\) indicates the maximum (supremum) of these intensities.
\end{definition}

If there is no path from \(u\) to \(v\), \(\widetilde{s}_d(u, v) = 0\).
If \(u = v\), \(\widetilde{s}_d(u, v) = 1\).
In UFSN, the connected intensity from \(u\) to \(v\) is equal to the intensity from \(v\) to \(u\). However, in DFSN, \(\widetilde{s}_d(u, v)\) is not necessarily equal to \(\widetilde{s}_d(v, u)\).

\begin{definition}
Assume \(G_{dn} = (V, E_{dn})\) is a directed fuzzy social network (DFSN). The directed fuzzy connected intensity matrix \(\widetilde{C}_d\) is a matrix where each element \(\widetilde{s}(v_i, v_j)\) represents the directed fuzzy connected intensity from node \(v_i\) to node \(v_j\):
    \[
    \widetilde{C}_d = \begin{pmatrix}
    \widetilde{s}(v_1, v_1) & \cdots & \widetilde{s}(v_1, v_n) \\
    \vdots & \ddots & \vdots \\
    \widetilde{s}(v_n, v_1) & \cdots & \widetilde{s}(v_n, v_n)
    \end{pmatrix}
    \]
\end{definition}

The directed fuzzy connected intensity matrix is vital for determining the relationships (direct or indirect) between any two actors in the DFSN. Hereafter, we will focus on TFNs using a the following compact notation: $\tilde{A}=(a_l,a_m,a_r).$

\section{Novel Fuzzy Centrality Indexes}

Centrality indices assess a node's importance in a network, reflecting the role of the corresponding entity in the modeled system.
In the following sections, we extend the classical centrality indexes to the case of a network where the edges are represented by triangular fuzzy numbers, i.e. ties are vague; thus, we define fuzzy in-degree centrality, fuzzy out-degree centrality, fuzzy total-degree centrality, fuzzy betweenness centrality,  fuzzy in-closeness centrality, fuzzy out-closeness centrality, and fuzzy total-closeness centrality.

\subsection{Fuzzy degree centrality}

In real-world interactions, we often consider people with many connections
to be important. Degree centrality transfers the same idea into a measure.
The degree centrality measure ranks nodes with more connections higher
in terms of centrality. In the classical framework, the degree centrality $C_d(v_i)$, in directed
graphs can be computed in different ways according to the number of in-connections and out-connections, leading to the in-degree index $d_i^{in}$ and the out-degree index $d_i^{out}$, respectively, or the combination of both\footnote{The number of edges connected to a vertex is referred to as its ``degree'' in graph theory. The degree of a node \( v_i \) is commonly denoted as \( d_i \). In the context of directed graphs, nodes have both in-degrees (edges pointing towards the node) and out-degrees (edges pointing away from the node), denoted as \( d_i^{in} \) and \( d_i^{out} \), respectively. For example, in social platforms like Facebook, the degree of a user corresponds to the number of friends \citep{Zafarani_2014}.
\begin{definition}
In any undirected graph \( G(V, E) \), the sum of all node degrees equals twice the number of edges:
\end{definition}
\begin{equation}
\sum_{i} d_{i} = 2|E|
\end{equation}
This equation holds because each edge contributes to the degree sum of exactly two nodes.
In directed graphs, the sum of in-degrees equals the sum of out-degrees:
\begin{equation}
\sum_{i} d_{i}^{in} = \sum_{j} d_{j}^{out}
\end{equation}
This equality arises because each directed edge contributes exactly one unit to the in-degree of its destination node and one unit to the out-degree of its origin node.
In real-world interactions, individuals with many connections are often perceived as important. Degree centrality quantifies this idea into a measurable index. The degree centrality index ranks nodes higher based on the number of connections they have.
For an undirected graph, the degree centrality \( C_{d} \) of a node \( v_i \) is given by:
\begin{equation}
C_{d}(v_i) = d_{i}
\end{equation}
where \( d_{i} \) represents the degree (number of adjacent edges) of node \( v_{i} \).
In directed graphs, we distinguish between in-degree centrality, out-degree centrality, and total degree centrality, which combines in and out degrees:
\begin{equation}
C_{d}(v_i) = d_{i}^{in} \quad \text{(In-degree centrality)}
\end{equation}
\begin{equation}
C_{d}(v_i) = d_{i}^{out} \quad \text{(Out-degree centrality)}
\end{equation}
\begin{equation}
C_{d}(v_i) = d_{i}^{in} + d_{i}^{out} \quad \text{(Total degree centrality)}
\end{equation}
In-degree centrality indicates how popular a node is within the network. Out-degree centrality measures how gregarious a node is.
Comparing degree centrality values across different networks directly is not straightforward due to variations in network size. To standardize comparison, degree centrality can be normalized as:
\begin{equation}
C_{d}^{*}(v_i) = \frac{d_i}{n-1}
\end{equation}
where \( n \) is the number of vertices in the network.
}.

However, in the context of (directed) fuzzy social networks, the edges are represented by fuzzy numbers, and thus they are not scalars.

\subsubsection{Fuzzy in-degree centrality index}

When using in-degrees, fuzzy degree centrality measures how popular a node is and its value shows prominence or prestige.

\begin{definition}
Assume that $\widetilde{G}_{dir}=\left(V, E, \widetilde{A}_{dir}\right)$ is a DFSN, $\widetilde{D}_{in}\left(v_{i}\right)$ is the sum of the fuzzy relations that are adjacent to $v_{i}$, then $\widetilde{D}_{in}\left(v_{i}\right)$ is called fuzzy in-degree centrality of $v_{i}$.
\end{definition}

For a given node $i$, let $\widetilde{O}_j^{in}$ be the set of fuzzy numbers $\widetilde{A}_{ji}$ ranked in decreasing order, indicating the relationship between each node $j$ and the node $i$ (directed), the formula of fuzzy in-degree centrality of $v_{i}$ is given by

\begin{equation}
\widetilde{D}_{in}\left(v_{i}\right)=\bigoplus_{j \in[n], j \neq i}\left(w_{j} \odot \tilde{O}_j^{in}\right)
\label{indegreefuzzy}
\end{equation}

\noindent where Equation \ref{indegreefuzzy} makes use of the FTOWA defined in Equation \ref{FTOWA}.
To rank fuzzy numbers, we use the centre of gravity approach\footnote{The Center of Gravity (CoG) of a fuzzy number \( \mu(x) \) is calculated as follows:
\begin{equation}
\operatorname{CoG}(\mu) = \frac{\int_{-\infty}^{\infty} x \cdot \mu(x) \, \mathrm{d}x}{\int_{-\infty}^{\infty} \mu(x) \, \mathrm{d}x}.
\label{CoG}
\end{equation}
\noindent where \( \mu(x) \) is the membership function of the FN \( \mu \), defined on the real numbers \( \mathbb{R} \) with values in the interval \([0,1]\); \( x \) represents the variable over which the integration is performed, encompassing all possible values within the support of the FN;
the numerator \( \int_{-\infty}^{\infty} x \cdot \mu(x) \, \mathrm{d}x \) computes the first moment (weighted average) of the FN, considering the distribution of \( \mu(x) \) over \( \mathbb{R} \), and finally, the denominator \( \int_{-\infty}^{\infty} \mu(x) \, \mathrm{d}x \) represents the total area under the membership function, which normalises the moment to yield the CoG.
The CoG is particularly useful in ranking FNs because it provides a real-valued metric that reflects the balance point of the FN. This method is especially effective when comparing TFNs, which are defined by their left endpoint \( a_l \), core \( a_m \), and right endpoint \( a_r \). The CoG value offers an intuitive means of comparing the magnitude of different FNs within the same context.
While other ranking methods exist, such as those based on medians or alternative moments, the CoG is favored in this study for its simplicity and alignment with the properties of TFNs.
}. It follows that we can obtain the following extreme cases:

\begin{itemize}
    \item $w_{j} = \{1,0, ... ,0\}$ $\implies \widetilde{D}_{in}$ is determined by the greatest $\widetilde{A}_{ji}$; 
    \item $w_{j} = \{0,0, ... ,1\}$  $\implies \widetilde{D}_{in}$ is determined by the lowest $\widetilde{A}_{ji}$; 
    \item $w_{j} = \{1/n, 1/n,..., 1/n\}$  $\implies \widetilde{D}_{in}$ is determined by all the $\widetilde{A}_{ji}$ with the same weights.  Therefore, we give the same importance to low and high fuzzy relationships in computing fuzzy in-degree centrality.
\end{itemize}

Equation \ref{indegreefuzzy} does not allow us to compare the centrality values between different networks. We can normalise the index to become a pure number between 0 and 1 to overcome this drawback. The reasoning behind the normalisation is that we calculate the maximum that the in-degree centrality index can take. The maximum of this index would be obtained if an actor is perfectly connected with all the other nodes. For this aim, the edge represented by the fuzzy number should have the characteristic of possessing a core with the maximum possible value on the support of the fuzzy number (with zero fuzziness). If the latter happens for all the actor's edges, we can compute the maximum of Equation \ref{indegreefuzzy}  as follows.

The easiest way to find this relative index is to normalise the fuzzy numbers. In fact, unlike the classical case in which the number of edges is necessary to normalise the in-degree centrality, in this case, all the nodes are connected, and therefore, the intensity of the edges matters. Once the fuzzy numbers have been normalised, we already have a relative index as follows:

\begin{equation}
\widetilde{D}^{in*}\left(v_{i}\right)=\bigoplus_{j \in[n], j \neq i}\left(w_{j} \odot \widetilde{O}_j^{in*}\right).
\label{indegreefuzzyrel}
\end{equation}

\noindent where $\widetilde{O}_j^{in*}$ is the order set of fuzzy numbers with normalised support in $[0,1]$.

\subsubsection{Fuzzy out-degree centrality index}

Out-degrees quantify the extent of a node's outgoing connections, measuring its influence or activity within the network. This metric reflects how actively a node interacts with others, indicating its level of engagement or gregariousness within the network's structure.

\begin{definition}
Assume that  $\widetilde{G}_{dir}=\left(V, E, \widetilde{A}_{dir}\right)$  is a DFSN, $\widetilde{D}^{out}\left(v_{i}\right)$ is the sum of the fuzzy relations that are adjacent from $v_{i}$, then $\widetilde{O}_j^{out}\left(v_{i}\right)$ is called fuzzy out-degree centrality of
$v_{i}$.
\end{definition}

For a given node $i$, let $\widetilde{O}_j^{out}$ be the set of fuzzy numbers $\widetilde{A}_{ij}$ ranked in decreasing order, indicating the relationship between each node $i$ and the node $j$ (directed), the formula of the fuzzy out-degree centrality of $v_{i}$ is given by:

\begin{equation}
\widetilde{D}^{out}\left(v_{i}\right)=\bigoplus_{j \in[n], j \neq i}\left(w_{j} \odot \tilde{O}_j^{out}\right).
\end{equation}

According to the vector $w_{j}$, we can obtain the following cases:

\begin{itemize}
    \item $w_{j} = \{1,0, ... ,0\}$ $\implies \widetilde{D}_{out}$ is determined by the greatest $\tilde{A}_{ij}$; 
    \item $w_{j} = \{0,0, ... ,1\}$  $\implies \widetilde{D}_{out}$ is determined by the lowest $\tilde{A}_{ij}$; 
    \item $w_{j} = \{1/n, 1/n,..., 1/n\}$  $\implies \widetilde{D}_{out}$ is determined by all the $\tilde{A}_{ij}$ with the same weights. Therefore, we give the same importance to low and high fuzzy relationships in computing fuzzy out-degree centrality.
\end{itemize}

A possible solution is to adopt a vector of decreasing weights so that the importance of fuzzy numbers decreases as the relationship is less important. As for the in-degree centrality index, also in this case, we can define a relative out-degree centrality index as follows:

\begin{equation}
\widetilde{D}^{out*}\left(v_{i}\right)=\bigoplus_{j \in[n], j \neq i}\left(w_{j} \odot \tilde{O}_j^{out*}\right).
\label{outdegreefuzzyrel}
\end{equation}

\subsubsection{Fuzzy total-degree centrality index}

In DFSN applications, the previous degrees can be of great interest. The fuzzy out-degree centralities are measures of expansiveness and the fuzzy in-degree centralities are measures of receptivity, or popularity. 

However, we can consider a formula for the fuzzy total-degree centrality of $v_{i}$ given by:

\begin{equation}
\widetilde{D}\left(v_{i}\right)=\widetilde{D}^{out}\left(v_{i}\right) \oplus  \widetilde{D}^{in}\left(v_{i}\right)
\end{equation}

\subsection{Fuzzy betweenness centrality}

Another way of looking at centrality is by considering how important nodes are in connecting other nodes. In the classical setting, for a node $v_i$, we can compute the relative frequency of the shortest paths\footnote{ 
A walk from node \( v_{0} \) to \( v_{k} \) in a graph \( G = (V, E) \) is an alternating sequence \( v_{0}, e_{1}, v_{1}, e_{2}, \ldots, e_{k}, v_{k} \) of vertices and edges, where \( e_{i} \) is \( \{v_{i-1}, v_{i}\} \) in the undirected case and \( (v_{i-1}, v_{i}) \) in the directed case, with braces indicating an undirected edge and parentheses representing an ordered pair for a directed edge.
The length of the walk is defined as the number of edges on the walk. The walk is called a path if \( e_{i} \neq e_{j} \) for \( i \neq j \), and a path is a simple path if \( v_{i} \neq v_{j} \) for \( i \neq j \). 
For a path \( p \) in a graph \( G = (V,E) \) with edge weights \( w \), the weight of the path, denoted by \( w(p) \), is defined as the sum of the weights of the edges on \( p \). A path from \( u \) to \( v \) in \( G \) is the shortest path (with respect to \( w \)) if its weight is the smallest possible among all paths from \( v_i \) to \( v_j \).} to connect other nodes that pass through $v_i$. The search for the shortest path consists of finding the path from one node to another that is as fast as possible, crossing the least number of intermediate nodes\footnote{For a node \( v_{i} \), one approach is to quantify the number of shortest paths between other nodes that pass through \( v_{i} \):
\begin{equation}
C_{b}(v_{i}) = \sum_{\substack{s \neq v_{i} \neq t}} \frac{\sigma_{st}(v_{i})}{\sigma_{st}}
\end{equation}
\noindent where \( \sigma_{st} \) denotes the total number of shortest paths from node \( s \) to \( t \) (also referred to as information pathways), and \( \sigma_{st}(v_{i}) \) represents the number of those paths that pass through \( v_{i} \). Essentially, this metric evaluates how central \( v_{i} \) is in facilitating connections between any pair of nodes \( s \) and \( t \).
To standardize betweenness centrality across networks, normalization is necessary as follows:
\begin{equation}
 C_{b}^{*}(v_{i}) = \frac{C_{b}(v_{i})}{\max_{s \neq t} C_{b}(v_{i})}   
\end{equation}
\noindent where the maximum value occurs when \( v_{i} \) lies on all shortest paths from \( s \) to \( t \) for every pair \( (s, t) \), i.e., \( \forall (s, t), s \neq t \neq v_{i}, \frac{\sigma_{st}(v_{i})}{\sigma_{st}} = 1 \).
}.

In fuzzy social networks, where fuzzy numbers represent connections between nodes, each node is connected to every other node with varying degrees of truth. As a result, if all fuzzy connections were equal, the shortest path between two nodes would be the direct connection between them. However, in real-world scenarios, the connections typically vary in strength, leading to different shortest paths depending on the specific degrees of truth associated with each connection.
Therefore, in our setting, we must reinterpret the concept of the shortest path as follows:  the importance of a node in connecting other nodes must be evaluated based on the quality of the existing fuzzy edges, and thus, one path is privileged over another if more effective in connecting two elements of the network.
In our context, a path can be considered more effective than another path if it can connect two nodes, ensuring a higher connection strength (it may happen that a longer path, in terms of intermediate nodes to cross, is better in terms of the strength of the edges and, therefore, more effective in connecting two elements $i$ and $j$). Hence, in a network where everyone is ``more'' or ``less'' connected, an indirect path to go from node $i$ to $j$ could be better than the directed path from  $i$ to $j$ (for example, $i$ and $j$ may have a weak direct connection but are strongly connected through a mutual acquaintance).

\begin{definition}
Given two different paths $P_{i \rightarrow j}^{(1)}$ and $P_{i \rightarrow j}^{(2)}$, of possible different lengths $k={1,2,..., K}$  and $t={1,2,...T}$, respectively, we
say that $P_{i \rightarrow j}^{(1)}$ is better than $P_{i \rightarrow j}^{(2)}$ when:

\begin{equation}
P_{i \rightarrow j}^{(1)}  \succ  P_{i \rightarrow j}^{(2)} \iff \bigoplus_{k \in [1,K]}\left(w_{k} \odot \tilde{N}_k^{(1)}\right) > \bigoplus_{t \in [1,T]}\left(w_{t} \odot \tilde{N}_t^{(2)}\right)
\label{eqq}
\end{equation}
\end{definition}

\noindent where $\widetilde{N}_k^{(1)}$ is the set of $K$ fuzzy edges in the first path between $i$ and $j$ with normalized support in $[0,1]$ and ranked in decreasing order, $\widetilde{N}_t^{(2)}$ is the set of fuzzy numbers in the second path between $i$ and $j$  also with normalized support in $[0,1]$ and ranked in decreasing order, and $w_{k}$ and $w_{t}$ are vectors of weights. We define $P_{i \rightarrow j}^{(1)}$ as the best path from $i$ to $j$ when for any other path $P_{i \rightarrow j}^{(2)}$ from $i$ to $j$ we have that $P_{i \rightarrow j}^{(1)}  \succ  P_{i \rightarrow j}^{(2)}$.
Equation \ref{eqq} allows us to consider different possibilities for selecting
the so-called best path, e.g. average fuzzy number, minimum, maximum, etc. From now on, we will use the minimum by introducing $w_{k}=\{0,0,...,1\}$ and $w_{t}=\{0,0,...,1\}$ as the vectors of weights. In other words, when looking at two possible alternative paths to reach $j$ starting from the node $i$, we examine all the fuzzy edges in those paths, and we select the best one with the higher minimum fuzzy number (edge) in the path.

\begin{figure}[htpb]
\centering
\includegraphics[width=8cm]{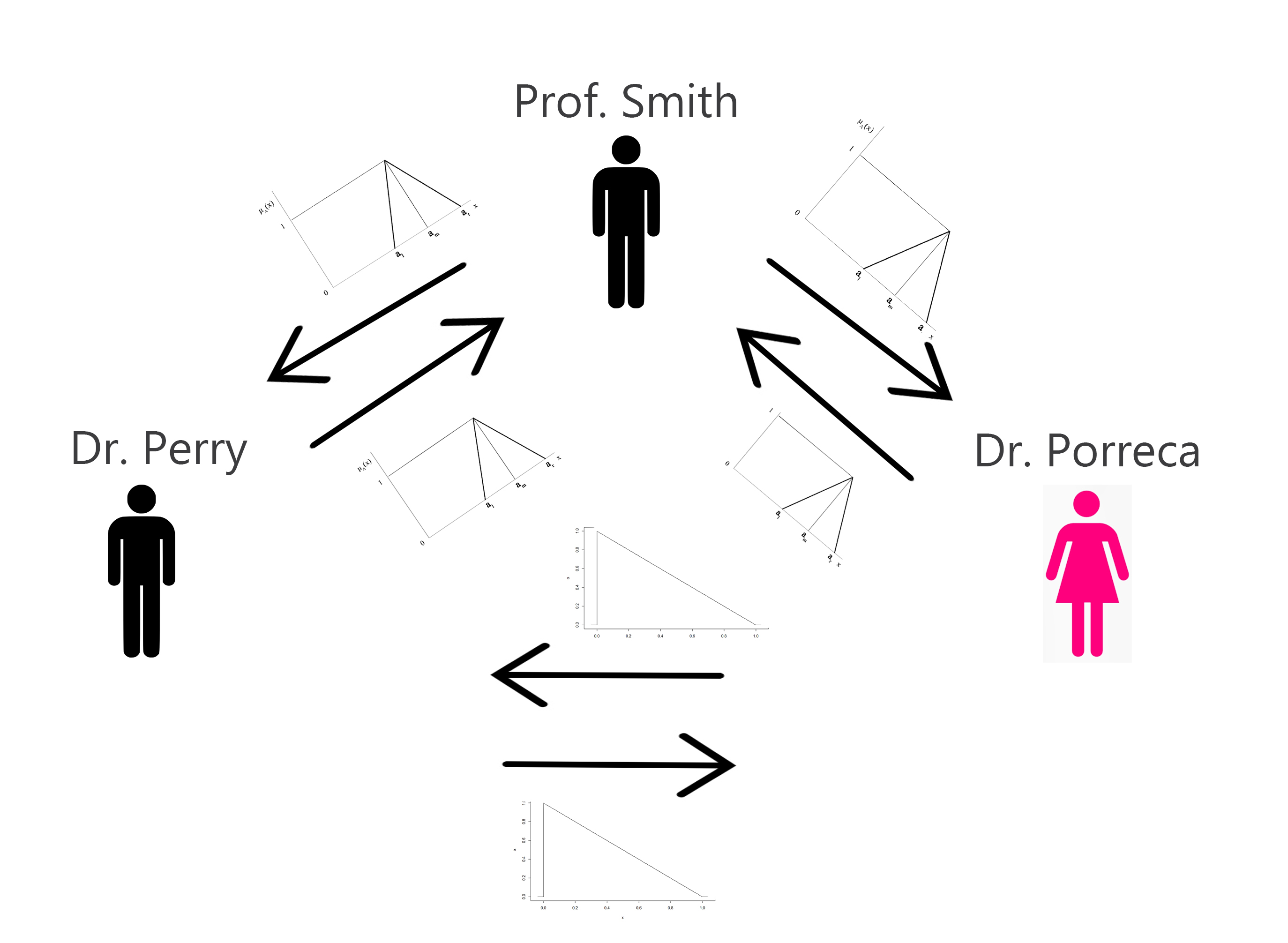}
\caption{Example of fuzzy social network where the directed path between two nodes is worse than an indirect path in terms of fuzzy numbers edges. In this context, the traditional concept of betweenness centrality in classic social networks, where an actor is considered highly central if they frequently lie on the shortest paths between two nodes, is redefined. Here, a node $z$ is considered highly central if it frequently appears on the optimal paths between any generic nodes $j$ and $i$.}
\label{esempioindiretto}
\end{figure}

An example that can help clarify the reasoning behind the proposed idea is illustrated in Figure \ref{esempioindiretto}. The best path from Dr. Perry to Dr. Porreca is via Prof. Smith. If we consider the direct connection, we have a very weak relationship, while if we connect the two nodes indirectly, we have two very strong relationships.
In other words, if the direct relationship is weak, it is better to take more steps characterised by strong relationships. In the case of the example shown, the same applies if Dr. Porreca wants to get in touch with Perry (reverse path). This graphic example helps to better understand that in our context, where the links are blurred, the concept of the shortest path is replaced by the idea of the "best" and most effective path to connect two distinct nodes.

Given the concept of the best path, we can define the fuzzy betweenness centrality as follows:

\begin{equation}
B\left(v_{i}\right)= \sum_{s \neq t \neq v_i} \frac{P_{st}(v_i)}{P_{st}}
\label{betfuzzy}
\end{equation}

\noindent where $P_{st}(v_i)$ indicate the number of ``best'' paths between $s$ and $t$ that passes thought $i$ (they are indirect paths), and $P_{st}$ is the total number of ``best'' paths between $s$ and $t$, i.e.  all the best paths between all the pairs of nodes $s$ and $t$ (direct or indirect paths via $i$ or not via $i$).

The search of the so-called ``best'' paths is done according to the minimum fuzzy number (edge) occurring in the path between $s$ and $t$, even if it is indirect. To compare the multiple different paths occurring in a social network, we rank them using the CoG (see Equation \ref{CoG}).

\subsection{Fuzzy closeness centrality}

In classical closeness centrality, the intuition is that the more central nodes, the more quickly they can reach other nodes. Formally, these nodes should have a smaller average shortest path length to other nodes. The smaller the average shortest path length, the higher the centrality for the node \footnote{Closeness centrality measures the average distance from a vertex to all other vertices in the network. Unlike other network metrics, closeness centrality offers a distinct perspective on each individual's network position, capturing how quickly each vertex can reach all others \citep{Hansen_2020}. The central idea is that nodes with higher closeness centrality are more central, allowing them to reach other nodes more efficiently \citep{Zafarani_2014}. Mathematically, closeness centrality \( C_{c}(v_{i}) \) for vertex \( v_{i} \) is defined as:
\begin{equation}
C_{c}(v_{i}) = \frac{1}{\bar{l}_{D_{j}}}
\end{equation}
\noindent where $\bar{l}_{D_{j}} = \frac{1}{n-1} \sum_{v_{j} \neq v_{i}} l_{i,j} $ is the average shortest path length from node \( v_{i} \) to all other nodes in the network. Nodes with smaller average shortest path lengths \( l_{i,j} \) exhibit higher closeness centrality}.

As we have observed with fuzzy betweenness centrality, the concept of the "shortest" path is replaced by the "best" path. This means that, when traveling from $i$ to $j$, one path might be preferable to another, even if it involves more intermediate nodes.

This reinterpretation naturally extends to fuzzy closeness centrality, where both in-closeness and out-closeness centrality are taken into account. Fuzzy in-closeness centrality reflects how central a node is based on its accessibility from other nodes in the network. A node with high fuzzy in-closeness centrality is characterized by strong incoming connections, indicating that it can be easily reached from various parts of the network.
On the other hand, fuzzy out-closeness centrality assesses a node's centrality based on its ability to connect to other nodes. A node with high fuzzy out-closeness centrality efficiently reaches many other nodes, underscoring its influence and reach within the network.
Thus, fuzzy closeness centrality provides a more nuanced understanding of a node's position, considering not just the shortest paths but the best paths. This approach allows for a richer analysis of centrality in networks where connections differ in strength and significance.

\subsubsection{Fuzzy in-closeness centrality}

Let the average (fuzzy number) of the best paths from node $i$ to all the nodes $j$ be represented by:

\begin{equation}
\widetilde{M}_{v_i, s} = \bigoplus_{v_j \neq v_i \in V} \left(w_{j} \odot \widetilde{A}_{ij}\right) 
\end{equation}

\noindent where $\widetilde{A}_{ij}$ is the set of the best fuzzy edges to connect $i$ and $j$ ranked in decreasing order, $s$ indicates the maximum number of steps considered evaluating the best path, and $w_{j}$ a vector of weights of the FTOWA operator. 

The fuzzy in-closeness centrality is defined as follows:

\begin{equation}
\widetilde{C}^{in}_{v_i,s} =\widetilde{M}_{v_i, s}
\label{closenessfuzzy}
\end{equation}

Thus, differently from Equation \ref{betfuzzy} that is a relative frequency (i.e. a crisp number), 
$\widetilde{C}^{in}_{v_i, s} $ is a centrality index that incorporates the fuzziness of the edges (i.e. a fuzzy number).

Another possible approach to define fuzzy in-closeness centrality, in a way that is more similar to the classical index, would be to consider $\widetilde{C^{(2)}}^{in}_{v_i, s} =\frac{1}{\widetilde{M}_{v_i, s}}$.
However, in the following, we refer to Equation \ref{closenessfuzzy} because our basic idea is to look for an immediate representation of the in-closeness concept rather than to identically  reproduce the classical indexes using fuzzy numbers. In fact, Equation \ref{closenessfuzzy} is easily interpretable, because provides a measure of how a node is central in considering the fuzzy incoming edges. However, Equation \ref{closenessfuzzy}, differently from \ref{indegreefuzzy} takes into account the concept of the best path and maximum number of steps to reach another node.

\subsubsection{Fuzzy out-closeness centrality}

Following the same reasoning, we can define the fuzzy out-closeness centrality index of the node $v_i$ as follows:

\begin{equation}
\widetilde{C}^{out}_{v_i,s} = \widetilde{Z}_{v_i, s}
\label{closenessfuzzyout}
\end{equation}

\noindent where $\widetilde{Z}_{v_i, s}$ is the average (fuzzy number) of the best paths from nodes $j$ to $i$ in decreasing order. Thus,

\begin{equation}
\widetilde{Z}_{v_i, s} = \bigoplus_{v_j \neq v_i \in V} \left(w_{j} \odot \widetilde{A}_{ji}\right) 
\label{closenessfuzzyoutmeans}
\end{equation}

\noindent is computed by applying the FTOWA operator to fuzzy numbers representing the edges from the generic nodes $j$ and the node $i$, for which we are computing the index.

As before, the vector of weights can be of different types. From now on, we will compute the average fuzzy number $\widetilde{M}_{v_i, s}$ or  $\widetilde{Z}_{v_i, s}$ using $w=\{1/n, ..., 1/n\}$. However, we stress that to select the best paths $P_{i \rightarrow j}$ or $P_{j \rightarrow i}$, we will always consider $w=\{0,...,1\}$.  Naturally, we could also consider different combinations of the weights according to the different purposes. 
As for the fuzzy in-closeness centrality index, also in this case we can consider an alternative measure of the fuzzy out-closeness centrality index as $\widetilde{C^{(2)}}^{out}_{v_i, s} =\frac{1}{\widetilde{Z}_{v_i, s}}$. This is more similar to the classical non-fuzzy in-closeness centrality index. The interpretation is contrary to \ref{closenessfuzzyout}; the lower the index, the higher the fuzzy in-closeness centrality index. Therefore, in the following, we focus on Equation  \ref{closenessfuzzyout} because the interpretation is quite immediate; indeed, the higher the index in Equation  \ref{closenessfuzzyout}, the higher the strength of the fuzzy outcoming edges of a node (given the number of steps to consider the best path), the higher the fuzzy out-closeness centrality of that node.

\subsubsection{Total fuzzy closeness centrality}

As we have seen for fuzzy degree centrality, we can also sum up the two different components of closeness centrality.
Therefore, we can define the total fuzzy closeness centrality index as follows:

\begin{equation}
\widetilde{C}^{tot}_{v_i, s} = _s\tilde{C}^{in}_{v_i, s} + _s\tilde{C}^{out}_{v_i, s} 
\label{closenessfuzzytot}
\end{equation}

\section{An application of fuzzy social network analysis to collaboration networks in University Departments}

Collaboration is a complex and multifaceted concept that cannot be evaluated using a single metric. Attempting to measure it directly takes a lot of work. One possible approach is to use proxies, such as counting the number of shared papers or similar information, but this method still needs to fully capture the essence of collaboration. Scientific collaboration is a broad concept involving exchanging knowledge and ideas between researchers. It includes sharing research procedures and ideas, which ultimately lead to the production of scientific knowledge. However, this definition alone is not enough to help us empirically measure a collaboration network of this type.

For this reason, this study starts from the basic idea of collecting data using the intuition of previous research in a different context. Notably, \citet{calcagni2014dynamic} proposed a new method to collect qualitative data using fuzzy set theory: the so-called \textit{Dynamic Fuzzy Rating Tracker} (DYFRAT). The authors developed an applet to collect directly triangular fuzzy numbers data without the fuzzy inference process \citep{calcagni2014dynamic}. Their work used mouse movements to model human rating evaluations from a fuzzy-set perspective, capturing the imprecision in point answers. In particular, DYFRAT captures the fuzziness of human ratings by modelling some real-time biometric events that occur during the cognitive rating process the noisy and dynamic x-y trajectory is initially mapped into a set of polar objects (distances and angles). Next, a linear histogram model of angles of movements is built, which stores the most important spatial features of the computer mouse trajectory, such as location, directions, and amplitudes. Finally, the histogram model is used to determine a quantification of spatial events involved in the original movement path.  DYFRACT  is not available, and thus, in this paper, following the intuition by \citep{calcagni2014dynamic}, we built an interactive fuzzy questionnaire to collect fuzzy numbers data to quantify and measure the collaboration relations between professors in the Economic Studies Department of the University ``\textit{G. d'Annunzio}'' in Italy. 

The software implemented to collect fuzzy data is based on the LAMP platform, i.e., \textit{Linux} (operating system) + \textit{Apache} (webserver) + \textit{MySql} (database management system) + \textit{PHP} (server-side programming language). For the management and use of the graphical interface was used Bootstrap (free framework initially created by Twitter to standardize graphical components, based in turn on \textit{HTML} + \textit{CSS} + \textit{JS}) with its many modules, while also retrieving the position of the mouse, were used: \textit{JS} (\textit{JavaScript}, object-oriented programming language and client-side events) + \textit{jQuery} (development library for client-side web applications based on JS).

\begin{figure}[htpb]
\centering
\includegraphics[width=13cm]{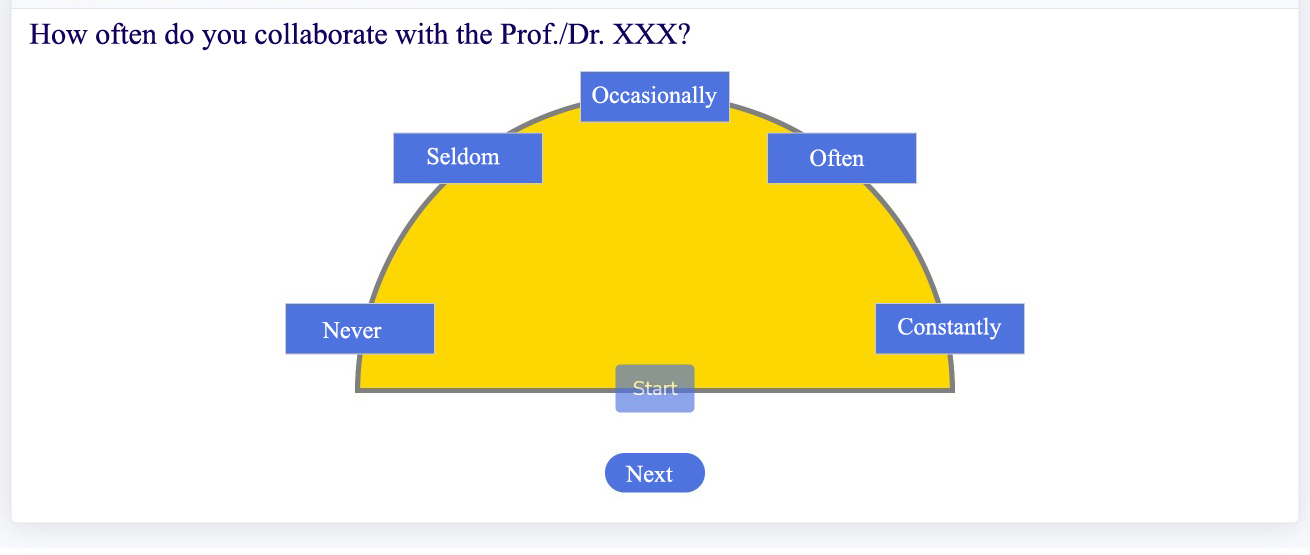}
\caption{Fuzzy questionnaire to collect mouse movements of respondents. \label{mov}}
\end{figure}

Figure \ref{mov} shows an example of a single questionnaire answer inserted on a semicircumference. Each user is called to answer the same question with respect to her/his collaboration relay with all the other members of the department.
This pseudo-circular scale is justified by its ability to provide greater degrees of freedom for mouse movement recordings than traditional linear or arc-type scales.
All the information related to the questions is saved directly on \textit{Database}. At the same time, the users' answers are stored on \textit{DB} in \textit{JSON} format (\textit{JS}-based information interchange language).

The distribution of mouse movements within the semicircumference, through the tracking procedure, allowed us to grasp the uncertainty in the answer before clicking on the final answer. This allowed us to build triangular distributions and reconstruct a triangular fuzzy number with its vagueness, support, core and membership function.
In other words, we were able to establish, for each professor, a relationship with all the other professors based on their subjective view of the relationship with their colleagues. This approach assumes the possibility that the relationship is asymmetric. Prof A might overestimate a collaboration with Prof. B, who, on the other hand, believes that the collaboration is of minor or significant importance.
Each professor answered the question for every possible department colleague, leading us to the creation of the entire departmental network.

\begin{figure}[htbp]
\centering
\includegraphics[width=10cm]{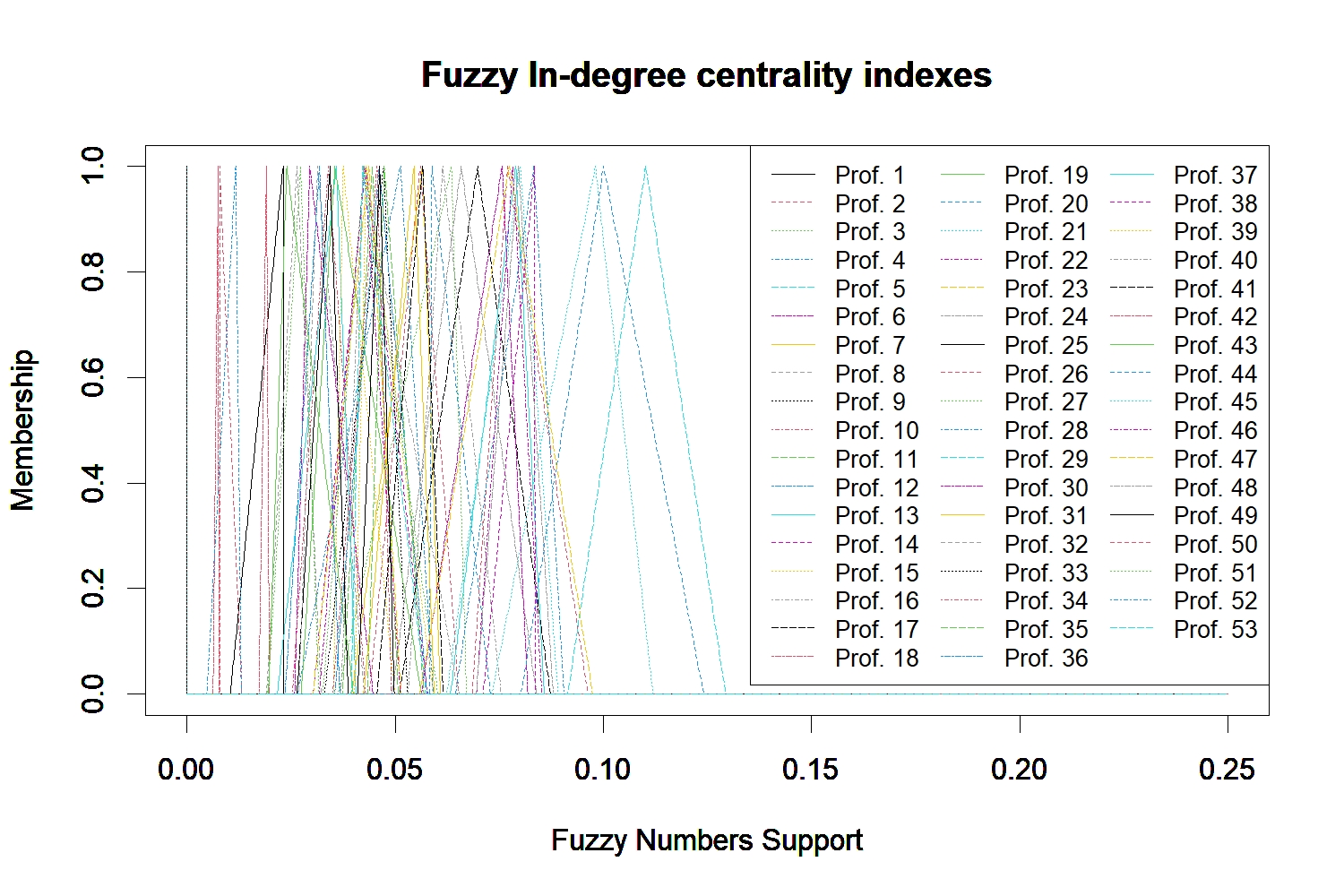}
\caption{Fuzzy in-degree centrality indexed of the actors in the collaborative network of the Department of Economic Studies of the University of Chieti-Pescara, Italy. } 
\label{dip2indegree}
\end{figure}

\begin{figure}
\centering
\includegraphics[width=10cm]{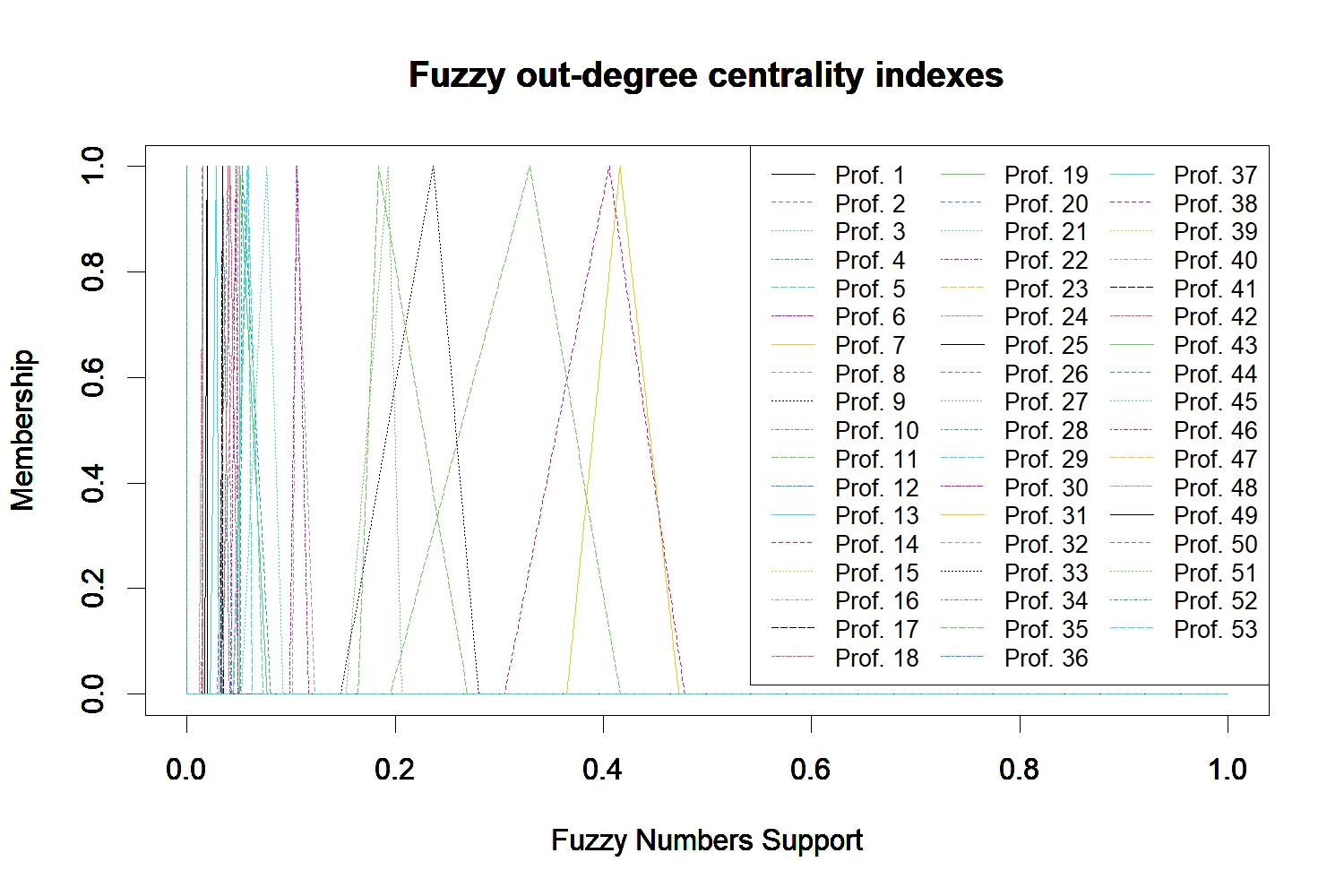}
\caption{Fuzzy out-degree centrality indexed of the actors in the collaborative network of the Department of Economic Studies of the University of Chieti-Pescara, Italy. } 
\label{dip2outdegree}
\end{figure}

Figure \ref{dip2indegree} and Table \ref{tab5} present the results of the fuzzy in-degree centrality indexes of the actors involved in the Department of Economic Studies of the University of Chieti-Pescara, Italy. The most central professors are 29, 44, and 21.  Clearly, for privacy reasons, we do not disclose the names of the department members.
The results can be evaluated both graphically and by observing the centre of gravity. Naturally, we could also consider other ranking methods among those available in the literature. 
Figure \ref{dip2outdegree} and Table \ref{tab6} represent the results of the fuzzy out-degree centrality indexes of the actors involved in the Department of Economic Studies of the University of Chieti-Pescara, Italy. The most central professors are 7, 38, and 35. It’s interesting to note how we get very different values if comparing in- and out-degree centrality indexes.

% latex table generated in R 4.0.2 by xtable 1.8-4 package
% Sun Jul 25 11:54:42 2021
\begin{table}[htpb]
\centering
\begin{tabular}{rrrrr}
  \hline
Prof  & Left & Core & Right & CoG \\ 
  \hline
29 & 0.09 & 0.11 & 0.13 & 0.11 \\ 
  44 & 0.08 & 0.10 & 0.12 & 0.10 \\ 
  21 & 0.07 & 0.10 & 0.11 & 0.09 \\ 
  26 & 0.07 & 0.08 & 0.10 & 0.08 \\ 
  28 & 0.06 & 0.08 & 0.09 & 0.08 \\ 
  37 & 0.06 & 0.08 & 0.09 & 0.08 \\ 
  38 & 0.07 & 0.08 & 0.08 & 0.08 \\ 
  45 & 0.06 & 0.08 & 0.09 & 0.08 \\ 
  46 & 0.07 & 0.08 & 0.09 & 0.08 \\ 
  47 & 0.06 & 0.08 & 0.10 & 0.08 \\ 
  48 & 0.07 & 0.08 & 0.09 & 0.08 \\ 
  17 & 0.05 & 0.07 & 0.09 & 0.07 \\ 
  24 & 0.05 & 0.07 & 0.08 & 0.07 \\ 
  30 & 0.06 & 0.08 & 0.08 & 0.07 \\ 
  2 & 0.04 & 0.06 & 0.07 & 0.06 \\ 
  3 & 0.04 & 0.06 & 0.07 & 0.06 \\ 
  16 & 0.05 & 0.06 & 0.08 & 0.06 \\ 
  20 & 0.06 & 0.06 & 0.07 & 0.06 \\ 
  4 & 0.03 & 0.05 & 0.06 & 0.05 \\ 
  7 & 0.04 & 0.06 & 0.06 & 0.05 \\ 
     \hline
\end{tabular}
\caption{First twenty Professors according to normalized Fuzzy in-degree centrality indexes in the collaborative network of the Department of Economic Studies of the University of Chieti-Pescara, Italy.}
\label{tab5}
\end{table}

\begin{table}[htpb]
\centering
\begin{tabular}{rrrrr}
  \hline
Prof & Left & Core & Right & CoG \\ 
  \hline
7 & 0.36 & 0.42 & 0.47 & 0.42 \\ 
  38 & 0.31 & 0.41 & 0.48 & 0.40 \\ 
  35 & 0.20 & 0.33 & 0.42 & 0.31 \\ 
  9 & 0.15 & 0.24 & 0.28 & 0.22 \\ 
  11 & 0.16 & 0.18 & 0.27 & 0.21 \\ 
  27 & 0.15 & 0.19 & 0.21 & 0.18 \\ 
  32 & 0.10 & 0.10 & 0.12 & 0.11 \\ 
  46 & 0.10 & 0.11 & 0.12 & 0.11 \\ 
  21 & 0.05 & 0.08 & 0.09 & 0.07 \\ 
  29 & 0.04 & 0.06 & 0.06 & 0.06 \\ 
  43 & 0.05 & 0.05 & 0.08 & 0.06 \\ 
  44 & 0.05 & 0.05 & 0.08 & 0.06 \\ 
  53 & 0.04 & 0.06 & 0.07 & 0.06 \\ 
  22 & 0.04 & 0.05 & 0.05 & 0.05 \\ 
  8 & 0.03 & 0.05 & 0.05 & 0.04 \\ 
  12 & 0.03 & 0.03 & 0.04 & 0.04 \\ 
  18 & 0.03 & 0.04 & 0.05 & 0.04 \\ 
  26 & 0.03 & 0.04 & 0.04 & 0.04 \\ 
  17 & 0.03 & 0.03 & 0.03 & 0.03 \\ 
  37 & 0.02 & 0.03 & 0.03 & 0.03 \\ 
     \hline
\end{tabular}
\caption{First twenty Professors according to normalized Fuzzy out-degree centrality indexes in the collaborative network of the Department of Economic Studies of the University of Chieti-Pescara, Italy.}
\label{tab6}
\end{table}

\begin{figure}[htpb]
\centering
\includegraphics[width=10cm]{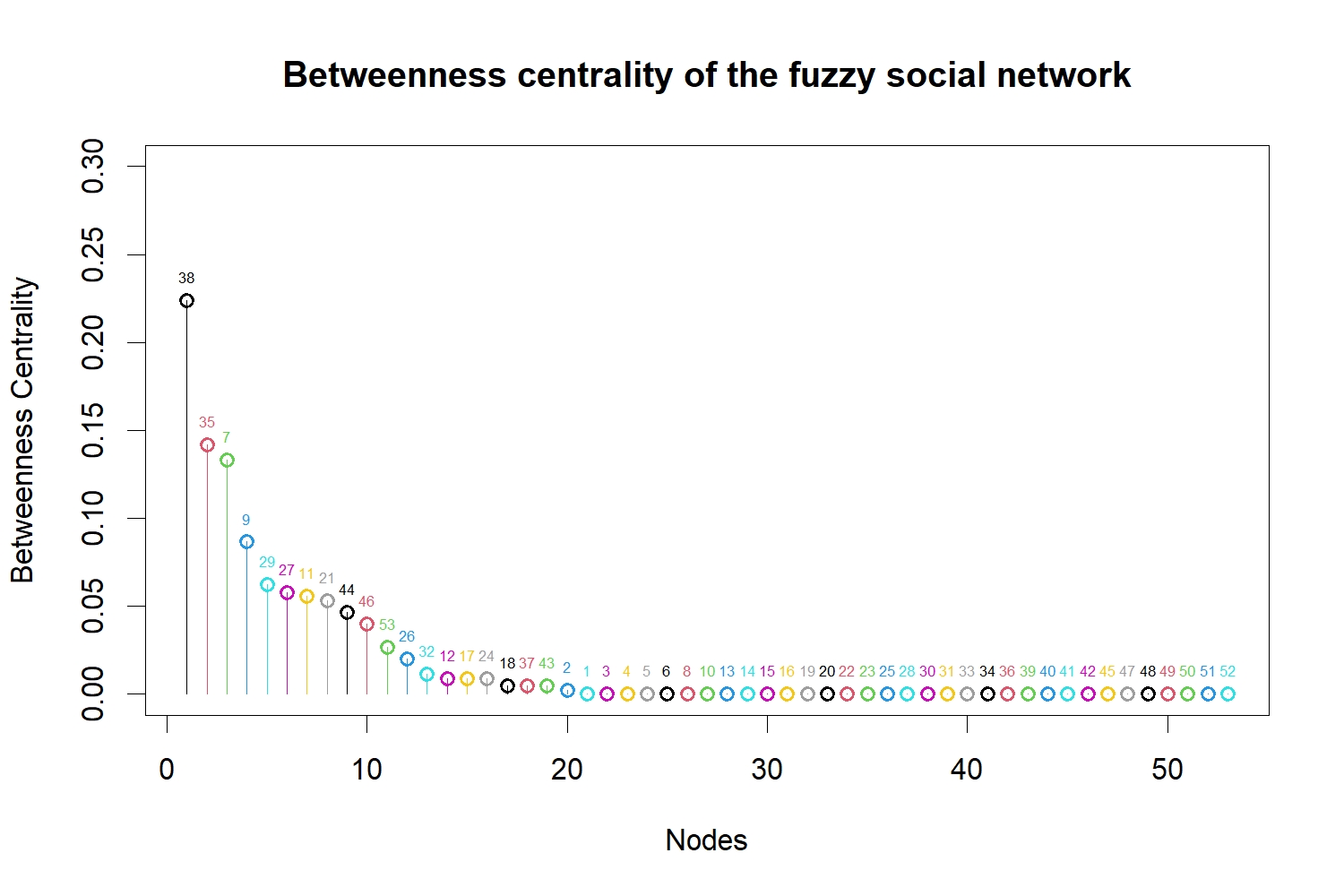}
\caption{Fuzzy betweenness centrality indexed of the actors in the collaborative network of the Department of Economic Studies of the University of Chieti-Pescara, Italy. } 
\label{dip2Betweenness}
\end{figure}

Figure \ref{dip2Betweenness} presents the findings of the fuzzy betweenness centrality indexes of the nodes of the Department of Economic Studies of the University of Chieti-Pescara, Italy. The most fundamental professors are 38, 35, and 7. In other words, professor 38 is often on the best path for many couple of professors because, for them, it's better to reach other colleagues passing via professor 38.

\begin{figure}[htpb]
\centering
\includegraphics[width=10cm]{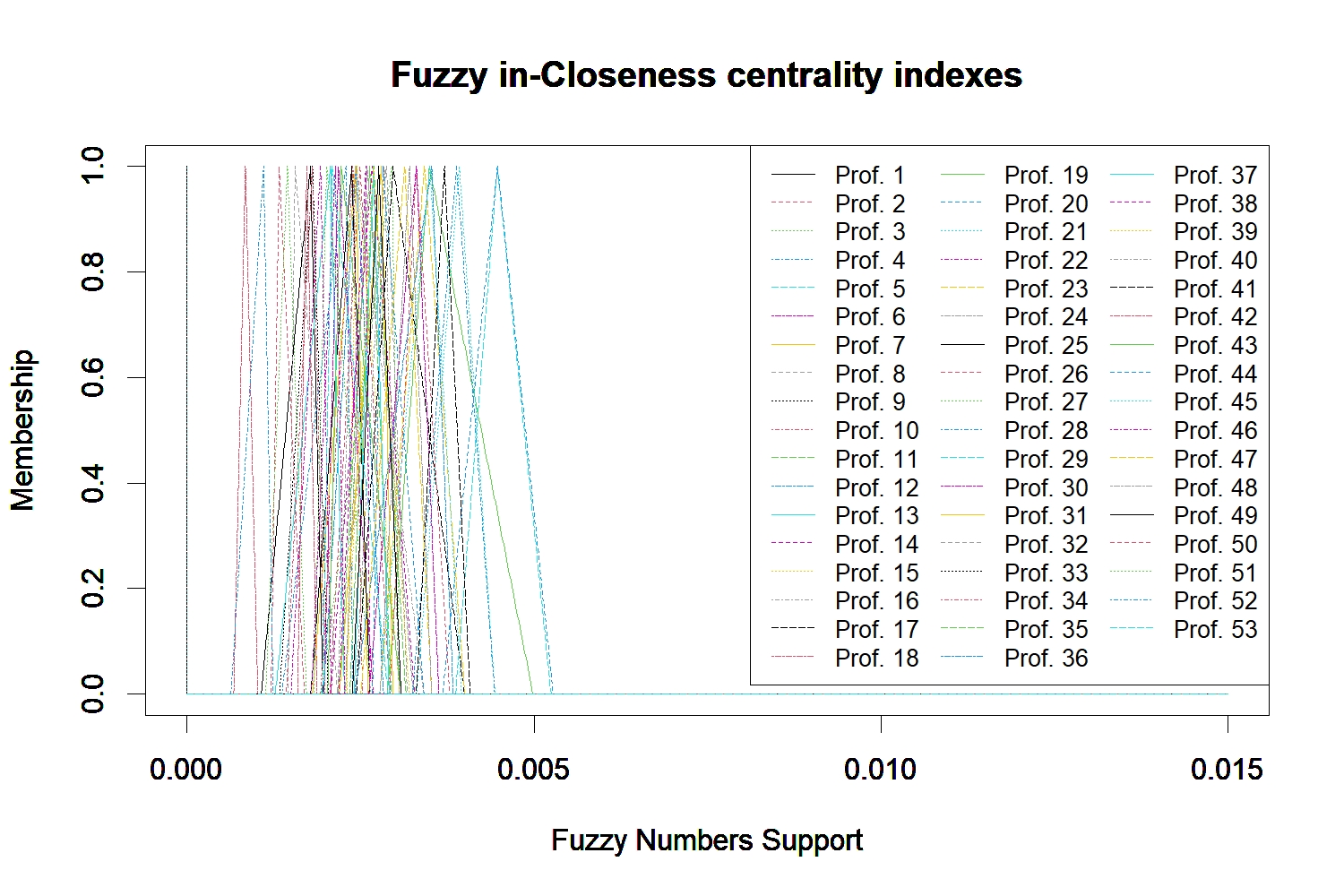}
\caption{Fuzzy in-closeness centrality indexed of the actors in the collaborative network of the Department of Economic Studies of the University of Chieti-Pescara, Italy. } 
\label{dip2incloseness}
\end{figure}

\begin{figure}[htpb]
\centering
\includegraphics[width=10cm]{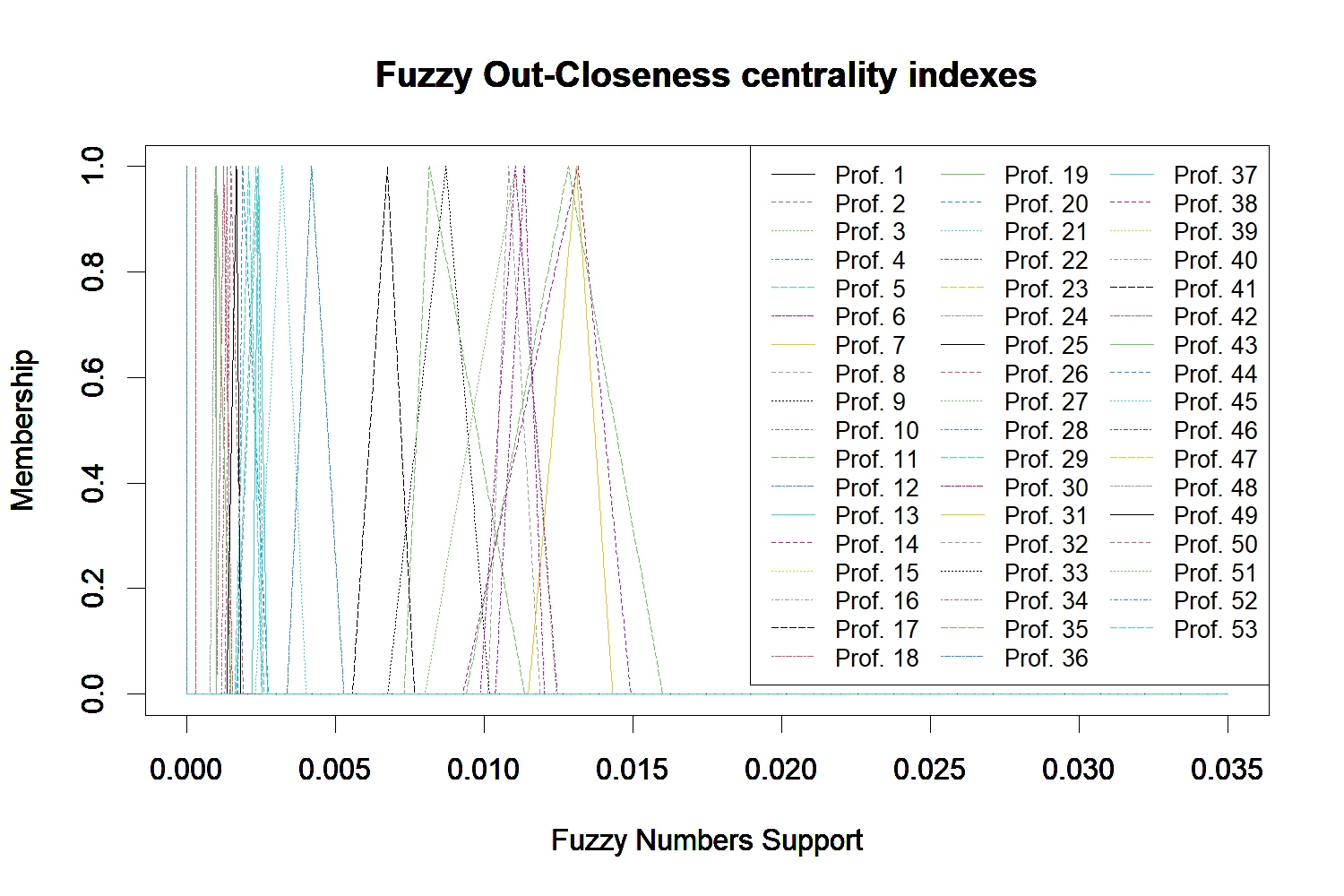}
\caption{Fuzzy out-closeness centrality indexed of the actors in the collaborative network of the Department of Economic Studies of the University of Chieti-Pescara, Italy. } 
\label{dip2outcloseness}
\end{figure}

Figure \ref{dip2incloseness},  Table \ref{tab7}, Figure \ref{dip2outcloseness},  Table \ref{tab8},  illustrate the results of the fuzzy in-closeness and out-closeness centrality indexes, respectively.
The most important nodes are 7, 35, 38, and 29, 44, 21, respectively.
In summary, professor 29 is the most powerful and approachable.

\begin{table}[htpb]
\centering
\begin{tabular}{rrrrr}
  \hline
Prof & Left & Core & Right & CoG \\ 
  \hline
7 & 0.0115 & 0.0131 & 0.0143 & 0.0130 \\ 
  35 & 0.0094 & 0.0128 & 0.0160 & 0.0127 \\ 
  38 & 0.0093 & 0.0131 & 0.0149 & 0.0124 \\ 
  22 & 0.0104 & 0.0113 & 0.0120 & 0.0112 \\ 
  46 & 0.0099 & 0.0110 & 0.0125 & 0.0111 \\ 
  32 & 0.0101 & 0.0108 & 0.0119 & 0.0109 \\ 
  27 & 0.0080 & 0.0110 & 0.0124 & 0.0105 \\ 
  11 & 0.0073 & 0.0081 & 0.0114 & 0.0089 \\ 
  9 & 0.0067 & 0.0087 & 0.0102 & 0.0085 \\ 
  17 & 0.0056 & 0.0067 & 0.0076 & 0.0066 \\ 
  12 & 0.0034 & 0.0042 & 0.0053 & 0.0043 \\ 
  21 & 0.0023 & 0.0032 & 0.0040 & 0.0032 \\ 
  37 & 0.0022 & 0.0024 & 0.0027 & 0.0024 \\ 
  8 & 0.0017 & 0.0024 & 0.0026 & 0.0022 \\ 
  29 & 0.0016 & 0.0023 & 0.0025 & 0.0022 \\ 
  44 & 0.0017 & 0.0019 & 0.0027 & 0.0021 \\ 
  53 & 0.0016 & 0.0021 & 0.0025 & 0.0021 \\ 
  49 & 0.0013 & 0.0017 & 0.0018 & 0.0016 \\ 
  26 & 0.0013 & 0.0015 & 0.0019 & 0.0015 \\ 
  10 & 0.0012 & 0.0013 & 0.0014 & 0.0013 \\ 
   \hline
\end{tabular}
\caption{First twenty Professors according to normalized Fuzzy in-closeness centrality indexes in the collaborative network of the Department of Economic Studies of the University of Chieti-Pescara, Italy.}
\label{tab7}
\end{table}

\begin{table}[htpb]
\centering
\begin{tabular}{rrrrr}
  \hline
 Prof & Left & Core & Right & CoG \\ 
  \hline
29 & 0.0039 & 0.0045 & 0.0052 & 0.0045 \\ 
  44 & 0.0037 & 0.0045 & 0.0053 & 0.0045 \\ 
  21 & 0.0033 & 0.0039 & 0.0044 & 0.0039 \\ 
  28 & 0.0033 & 0.0039 & 0.0044 & 0.0039 \\ 
  19 & 0.0029 & 0.0035 & 0.0050 & 0.0038 \\ 
  41 & 0.0033 & 0.0037 & 0.0041 & 0.0037 \\ 
  45 & 0.0028 & 0.0035 & 0.0040 & 0.0034 \\ 
  23 & 0.0028 & 0.0034 & 0.0040 & 0.0034 \\ 
  26 & 0.0029 & 0.0033 & 0.0038 & 0.0033 \\ 
  36 & 0.0024 & 0.0035 & 0.0038 & 0.0033 \\ 
  30 & 0.0026 & 0.0033 & 0.0036 & 0.0032 \\ 
  48 & 0.0028 & 0.0032 & 0.0035 & 0.0032 \\ 
  17 & 0.0024 & 0.0030 & 0.0040 & 0.0031 \\ 
  47 & 0.0025 & 0.0031 & 0.0035 & 0.0031 \\ 
  20 & 0.0027 & 0.0028 & 0.0034 & 0.0030 \\ 
  16 & 0.0024 & 0.0028 & 0.0031 & 0.0028 \\ 
  49 & 0.0024 & 0.0028 & 0.0031 & 0.0027 \\ 
  8 & 0.0022 & 0.0029 & 0.0032 & 0.0027 \\ 
  35 & 0.0023 & 0.0027 & 0.0032 & 0.0027 \\ 
  39 & 0.0023 & 0.0027 & 0.0032 & 0.0027 \\ 
   \hline
\end{tabular}
\caption{First twenty Professors according to normalized Fuzzy out-closeness centrality indexes in the collaborative network of the Department of Economic Studies of the University of Chieti-Pescara, Italy.}
\label{tab8}
\end{table}

\section{Discussion e conclusions}

In the literature on social network analysis, the links are binary, i.e. present or absent. A possible extension of this approach, widely used in the literature, considers the so-called weighted networks. In the latter, the link between the different nodes of a network is generally expressed with a number in the range $[0, 1]$. 
This approach of considering connections with a weight does not align with fuzzy logic's philosophy. Fuzzy logic embraces the coexistence of opposites and, for this reason, typically conveys information about vagueness. Attempting to translate an inherently imprecise human term into a scalar sacrifices information for simplicity.

This work considers the possibility that a vague measure can express the link between different actors in a social network. 
This approximate measure could be due to an attribute of human language that is  inaccurate by nature; for example, the inaccuracy inherent in concepts such as being friends, being little friends, being very close friends, being close collaborators, or having a poor collaboration, etc. In all these cases, we should have imprecise edges to preserve information about the vagueness of these links. 
In other cases, as in the application proposed in this work, we can also experience another kind of uncertainty due to indecision in choosing the final answer to a questionnaire composed by an ordinal variable, for example, by checking the mouse's movements before providing the definitive answer. Trying to preserve information about the vagueness of relationships within centrality indices, although the concepts of variability and vagueness are quite different, has something in common with the importance in statistics of preserving information about the variability of the distribution; in other words, it provides us with much richer information than a simple crisp link.

The translation of these sources of vagueness has been addressed using triangular fuzzy numbers. This approach extends classic social networks to scenarios where actors are fully connected through nuanced ties, represented by functions. Consequently, this paper introduces a novel method for analyzing social networks, where the connections between actors are expressed through fuzzy numbers. In other words, the links between nodes are represented by functions rather than scalars, as in traditional social network theory.
In light of this new context, this study presented an extension of the traditional centrality indices to the case where the network consists of shaded edges. This research introduces several new measures that consider the imprecision of the links in calculating the centrality measures and, in particular, often provides an original interpretation of these new measures. 

In the application part of this work, we presented a case of a collaboration network within a university department. The goal was to measure individual collaboration through questionnaires to understand how much each pair of professors or researchers collaborates. The basic idea of this approach is to try to understand the links between department members that go beyond simple indicators such as the number of publications in common or other parameters that can be measured through scalar measures. Indeed, in a department or even in another type of network, there are many cases in which the friendship or, more generally, the collaboration between two members goes beyond the number of tangible products resulting from this collaboration. Strong cooperation between individuals can also exist because they have projects in common, exchange ideas, exchange didactic material, socialize together outside of the workplace, are relatives, and help each other with research questions even if they are not published together, have interests in common outside the university for example for professional activities, or different types of interests in common. In all these cases, considering objective parameters such as the number of joint publications would be highly limiting. In any case, since many of these concepts we have listed are by their nature vague, the relationship between individuals and, therefore also, their centrality in a more complex network must be measured through suitable methods that consider the imprecision of human language as well as the uncertainty connected with human relationships. For the latter reason, the social network we propose in this research is rarely symmetrical. In fact, in most cases, an individual's opinion regarding his/her relationship with another individual is rarely the same in the opposite direction. For this reason, we have focused on the case of direct social networks. 

An exciting research component is the possibility of choosing a system of different weights to calculate the proposed centrality indices. Indeed, in some specific contexts and depending on the research objective, it could be interesting to consider other systems of weights.
In the theoretical part of the research, we introduced generalized measures to allow the calculation of different centrality measures based on different possible systems of weights. In the application part, there were no particular contextual requirements to think of weights other than those all equal to each other, so we focused on a system of equal weights.

In attributing weights to calculate the centrality indices, choosing the method to create a ranking of fuzzy numbers also assumes great importance. Different methods of sorting fuzzy numbers could lead to different results. This work focused on the centre of gravity method, but our approach can be extended to other procedures. The research focused on triangular fuzzy numbers, but of course, a possible exciting extension would be to consider different membership functions to represent the links existing in a social network.

Our approach's main advantage is to analyse networks of actors in which the relationships are not precise because they are defined by heterogeneous components or simply because they are collected through questionnaires whose possible answers are imprecise or simply attributes of human language. In this research, we presented an application of the data collected within university departments, but this approach can be extended to multiple real contexts. For example, we could consider other workplaces or all networks of individuals involved in sports or government teams. Consequently, the practical applications are numerous and of considerable interest.

\section*{Competing interests}
The authors declare that they have no known competing financial interests or personal relationships that could have appeared to influence the work reported in this paper.

% \section*{Author contributions statement}
% A.P. conceived and conducted the survey. A.P and F.M. developed the methodology, made the statistical analysed of the collected data and contributed to writing the manuscript. V.V. contributed to writing the manuscript and checking the mathematical parts.

\section*{Acknowledgments}
The authors thank Dr. Simone Di Nardo for contributing to the data collection platform and professors Marta Di Nicola and Francesca Scozzari for supporting Dr. Annamaria Porreca in her doctoral path, during which she collected data and worked on the project. We also thank the members of the Department of Economics of the University of Pescara-Chieti for responding to the questionnaire and allowing the study to be carried out.

\bibliography{_biblio}

\end{document}